\documentstyle[12pt,aps,epsfig]{revtex}

\title{\bf \vspace{5cm}
Resonant diffraction radiation from an ultrarelativistic
particle moving close to a tilted grating}
\author{A. P. Potylitsyn,
P. V. Karataev \thanks{corresponding author: e-mail: kpv@npi.tpu.ru}, and
 G. A. Naumenko}

\begin{document}

\maketitle

\begin{center}
Institute for Nuclear Physics

Tomsk Polytechnic University,

634050, pr.Lenina 2A, Tomsk, Russia

\end{center}

\begin{abstract}
A simple model for calculating the diffraction radiation characteristics
from an ultrarelativistic charged particle moving close to a tilted ideally
conducting strip is developed. Resonant diffraction radiation (RDR) is
treated as a superposition of the radiation fields for periodically spaced strips.
The RDR characteristics have been calculated as a function of the number of grating
elements, tilted angle, and initial particle energy. An
analogy with both the resonant transition radiation in absorbing medium and
the parametric X-ray radiation is noted.
\end{abstract}

\begin{center}
\section*{Introduction}
\end{center}

To date, a number of different approaches [1-4] have been employed to treat the
characteristics of electromagnetic radiation from a
relativistic particle moving parallel to and above diffraction grating (so-called
Smith-Purcell effect (SPE)). The interest in this type of radiation
is related to both a possibility of using the SPE
to generate intense radiation in the millimeter and submillimeter
regions [5] and to implement it in non-destructive beam diagnostics
[6]. In both cases we have to estimate the influence of such factors
as transverse beam size, angular beam divergence, monochromaticity, etc.
on the radiation characteristics.

In principle, the beam divergence effects in a plane that is parallel to the
grating can be estimated using the results obtained in [4]. However, there is
no simple algorithm to calculate the radiation characteristics from a
particle passing above a tilted grating, i.e. for a non-zero angle between the particle
trajectory and the main plane of the grating.

One of the authors of the present paper has developed an
approach based on description of SPE as a resonant diffraction radiation (RDR) [2]
which is suitable for calculating the radiation characteristics for a grating
consisting of a number of conducting strips spaced by vacuum
gaps. Here we calculate the RDR characteristics
of a particle whose trajectory is not parallel to the grating.
 Also, the influence of geometry on the RDR characteristics is
 studied.

\hspace{1cm}\section*{Diffraction radiation (DR) from an ultrarelativistic particle for a tilted strip}

To calculate the RDR characteristics for a tilted grating it is
neccessary to know the diffraction radiation field from a charged particle moving
close to a single tilted strip.

The exact solution of Maxwell equations
describing the radiation from a charged particle moving above an inclined semi-infinite
ideally conducting screen has long been known [7]. Using the latter and the results reported in [8],
we can obtain an expression for the DR field strength for a tilted strip  as
a difference between the radiation fields of two semi-infinite planes, one restricted by edge 1,
and the other $-$ by  edge 2 (see Fig.1)

\begin{equation}
\stackrel{\rightarrow}{E}_{strip}=\stackrel{\rightarrow}{E}_{up}
	- \stackrel{\rightarrow}{E}_{down}
\end{equation}

For convinience, we express the impact parameter (the shortest distance
between the particle trajectory and the plane edge) for edge 1 in the following
way:

\begin{equation}
a_1=h-\frac{a}{2} \sin\Theta_0,
\end{equation}

whereas for edge 2

\begin{equation}
a_1=h+\frac{a}{2} \sin\Theta_0.
\end{equation}

Here $h$ is the spacing between the particle tragectory and the middle line of the strip,
$a$ is the strip width, and $\Theta_0$ is the strip tilt angle. Thus, taking into account the
phase shift we have

\begin{equation}
\stackrel{\rightarrow}{E}_{strip}=\stackrel{\rightarrow}{E}_{DR}
(h-\frac{a}{2} \sin\Theta_0)e^{i\phi}-\stackrel{\rightarrow}{E}_{DR}
(h+\frac{a}{2} \sin\Theta_0)e^{-i\phi}
\end{equation}

In (4) the DR field for a semi-infinite ideal screen is expressed through
$\stackrel{\rightarrow}{E}_{DR}(a_i)$. The full phase shift ($2\phi$)
characterizes the phase difference between the waves being formed in the vicinity of
edges $1$ and $2$ [9]; and it can be derived from simple geometrical relations as a
quantity which is propotional to the time difference between the wave propagation from edge $1$
and edge $2$ (see Fig.1)

\begin{equation}
2\phi=\frac{2 a\pi\left [\cos(\Theta_y-\Theta_0)-\frac{\cos\Theta_0}{\beta}\right]}{\lambda},
\end{equation}

where $\beta$ is the particle velocity, and $\lambda$ is the DR wavelength.

In this paper use is made of the system of units $h=m=c=1$.

For the extreme cases ($\Theta_0=0^0$ and $90^0$), the phase shift calculated
according to (5) coincides with that obtained in [2].

Equation (4) is valid if the following conditions for the strip dimensions
are fulfilled:
$\gamma\lambda>>b, \gamma\lambda<<c$ (see Fig.1). Here $\gamma$ is the Lorentz -
factor of the particle.

We will consider the radiation characteristics in a coordinate system
where the $z$ axis is directed along the beam, the $x$ axis is parallel to and
the $y$ axis is perpendicular to the strip edge. As shown in [8], for relativistic
particles the radiation concentrates in the range of angles

\begin{equation}
\vert\Theta_x\vert\leq\gamma^{-1},
\end{equation}

if $\gamma>>1$.

In this approximation the following expression for the DR yield with
wavelength $\lambda$ will be true:

\begin{equation}
\stackrel{\rightarrow}{E}_{DR}(x + \Delta x)=\stackrel{\rightarrow}{E}_{DR}(x)
\exp\left [-\frac{2\pi\Delta x}{\gamma\lambda}\sqrt{1+\gamma^2\Theta^2_x}\right],
\end{equation}

where $\stackrel{\rightarrow}{E}_{DR}(x) \sim \exp\left\{ -\frac{\omega}{2\omega_c}
\sqrt{1+\gamma^2\Theta^2_x}\right\}$, and $\omega_c=\frac{\gamma}{2a_1}$ is the
DR characteristic energy.

For further calculations we shall use a more symmetrical expression instead of (4)

\begin{equation}
\stackrel{\rightarrow}{E}_{strip}=\stackrel{\rightarrow}{E}_{DR}(h)
\left [\exp(\alpha+i\phi)-\exp(-\alpha-i\phi)\right],
\end{equation}

\begin{equation}
\alpha=\left (\frac{a \pi \sin\Theta_0}{\gamma\lambda}\right )\sqrt{1+\gamma^2\Theta^2_x}
\end{equation}

From (8) we derive the following fomula for the DR spectral-angular density  for the
strip:

\begin{equation}
\frac{d^2 W_{strip}}{d\omega d\Omega}=\frac{d^2 W_{DR}}{d\omega d\Omega}
F_{str},
\end{equation}

\begin{equation}
\frac{d^2 W_{DR}}{d\omega d\Omega}=4\pi^2|\stackrel{\rightarrow}{E}_{DR}|^2,
 F_{str}=4\left(\sinh^2\alpha+\sin^2\phi\right)
\end{equation}

The expressions obtained are quite similar to the formulas for the spectral - angular
distribution of transition radiation (TR) from a foil (see, e.g., [10]).
In the case in question, $F_{str}$ characterizes the DR field interference from
the two strip edges;
whereas, for TR, the same multiplier characterizes the TR
field interference from the input and output surfaces of the foil.

Earlier [9], the spectral-angular density of DR for a semi-infinite screen has been obtained
using an ultrarelativistic approximation. It was shown that the DR is
concentrated in the vicinity of the plane that is perpendicular to the
screen ($\Theta_x \sim \gamma^{-1}$)

\begin{eqnarray}
&  \frac{d^2 W_{DR}}{d\omega d\Omega}=\frac{\alpha}{4\pi^2}\exp\left(-\frac{\omega}
{\omega_c}\sqrt{1+\gamma^2\Theta^2_x}\right)\times \nonumber\\
&  \times\{\Theta^2_x[1+\cos(\Theta_y-\Theta_0)](1-\cos\Theta_0)+ \nonumber\\
&  (\gamma^{-2}+\Theta^2_x)[1-\cos(\Theta_y-\Theta_0)](1+\cos\Theta_0)\}\times \nonumber\\
&  \times\{(\gamma^{-2}+\Theta^2_x)\{[\cos(\Theta_y-\Theta_0)-\frac{\cos\Theta_0}{\beta}]^2+ \nonumber\\
&  +(\gamma^{-2}+\Theta^2_x)\sin^2\Theta_0\}\}^{-1}
\end{eqnarray}

In (12), we omitted the terms smaller than $\sim\gamma^{-2}$
in the numerator and denominator the terms.

Now we shall consider the forward diffraction
radiation (FDR), i.e., for the angles $\Theta_y\sim\gamma^{-1}<<1$.
Using this approximation instead of (12) we have

\begin{equation}
\frac{d^2 W_{DR}}{d\omega d\Omega}=\frac{\alpha}{4\pi^2}\exp\left(-\frac{\omega}
{\omega_c}\sqrt{1+\gamma^2\Theta^2_x}\right)\frac{\gamma^{-2}+2\Theta^2_x}
{(\gamma^{-2}+\Theta^2_x)(\gamma^{-2}+\Theta^2_x+\Theta^2_y)}
\end{equation}

As was noted in [8], in the angular distribution of DR at $\Theta_y=
\Theta_x=0$ there is a maximum whose value is proportional to $\gamma^2$

\begin{equation}
\frac{d^2 W_{DR}(\Theta_x=0,\Theta_y=0)}{d\omega d\Omega}=
\frac{\alpha}{4\pi^2}\gamma^2
\exp\left(-\frac{\omega}{\omega_c}\right)
\end{equation}

For $\Theta_y>>\gamma^{-1}$, using (12) one can obtain a simpler
expression:

\begin{eqnarray}
& \frac{d^2 W_{DR}}{d\omega d\Omega}=\frac{\alpha}{4\pi^2}
\exp\left(-\frac{\omega}{\omega_c}\sqrt{1+\gamma^2\Theta^2_x}\right)\times \nonumber\\
& \times\{\gamma^{-2}(1+\cos\Theta_0)[1-\cos(\Theta_y-\Theta_0)]+ \nonumber\\
& +2\Theta^2_x
[1-\cos\Theta_0\cos(\Theta_y-\Theta_0)]\}\times \nonumber\\
& \times\{(\gamma^{-2}+\Theta^2_x)\sin^2\frac{\Theta_y}{2}\sin^2\left(\Theta_0-\frac{\Theta_y}
{2}\right)\}^{-1}
\end{eqnarray}

As follows from (12), for the mirror reflection angle ($\Theta_y=2\Theta_0$) the
angular distribution has another maximum which is much the same as Eq. (14) and can be
identified as "backward diffraction radiation" (BDR) in analogy to the process of transition radiation.
The fact that the FDR and BDR intensities coincide in all the
frequency range results from using an ideally conducting screen
approximation. It is obvious that, however, the latter approximation is not valid for $\omega_c\geq\omega_p$,
where $\omega_p$ is the plasmon energy of the screen material.

In the range of angles $\vert2\Theta_0-\Theta_y\vert\sim\gamma^{-1}$, i.e., in
the vicinity of the mirror reflection direction, the BDR spectral-angular density
also has the form (13); however, in this case the angle $\Theta_y$ is
measured from the mirror reflection direction (see [8]).

Thus, expression (15) is valid when the following conditions are fulfilled:

\begin{equation}
\Theta_y>>\gamma^{-1},\hspace{6mm} \vert2\Theta_0-\Theta_y\vert>>\gamma^{-1}
\end{equation}

In this case, as follows from (15), the DR density is about $\gamma^2$ times smaller.

 Figure 2 shows the dependence of the DR yield	on the strip tilt angle $\Theta_0$
(so-called orientation dependence) for the fixed observation angle $\Theta_y=4.5^0$
in the reflection plane ($\Theta_x=0$). The calculations have been carried out
using the formulas (10)$-$(12) for $\gamma=1000$, $\lambda=0.4 \mu m$, and $a_1=0.1mm$.
One can notice that for $\gamma\lambda\geq a\sin\Theta_0$ the DR yield
is strongly suppressed and the characteristic angular width of the dependence
is significantly higher than $\sim\gamma^{-1}$ which is typical for
DR from a semi-infinite screen (see Fig.2d). The DR intensity at the orientation
dependence maximum is also suppressed when $\gamma\lambda\geq a\sin\Theta_0$.

\begin{center}
\section*{Resonant diffraction radiation from a tilted grating}
\end{center}

Let us consider a grating consisting of $N$ strips, of width $a$
and period $d$ tilted at angle $\Theta_0$ to the electron momentum
(see Fig.3). The impact parameter (the distance between the first strip center and the electron
trajectory (1)) we denote as $h$.

The radiation field being formed near strip $1$ of the grating coincides with expression
(8)

\begin{equation}
\stackrel{\rightarrow}{E}_{1}=\stackrel{\rightarrow}{E}_{str}(h)
\end{equation}

whereas the next strip field differs from (17) by both the phase $\phi_0$
and the decay factor $\alpha_0$

\begin{equation}
\stackrel{\rightarrow}{E}_{2}=\stackrel{\rightarrow}{E}_{1}\exp(-\alpha_0-i\phi_0)
=\stackrel{\rightarrow}{E}_{str}(h)\exp(-\alpha_0-i\phi_0),
\end{equation}

which can be determined in analogy to (5) and (9)

\begin{eqnarray}
\phi_0=\frac{2\pi d\left [\cos(\Theta_y-\Theta_0)-\frac{\cos\Theta_0}{\beta}\right]}{\lambda}, \\
\vspace{12mm}
\alpha_0=\left (\frac{2\pi d\sin\Theta_0}{\gamma\lambda}\right )\sqrt{1+\gamma^2\Theta^2_x}
\end{eqnarray}

One can write the $k^{th}$ strip field in the same way

\begin{equation}
\stackrel{\rightarrow}{E}_{k}=\stackrel{\rightarrow}{E}_{str}
\exp[-(k-1)(\alpha_0+i\phi_0)]
\end{equation}

The resulting field  of the $N$ strip grating is expressed through the sum of $N$
terms

\begin{equation}
\stackrel{\rightarrow}{E}_{GR}=\stackrel{\rightarrow}{E}_{1}+
\stackrel{\rightarrow}{E}_{2}+\dots+\stackrel{\rightarrow}{E}_{k}=
\stackrel{\rightarrow}{E}_{str}\sum\limits^N_{k=1}\exp[-(k-1)(\alpha_0+i\phi_0)]
\end{equation}

Having calculated the squared modulus of expression (22) we obtain the following
expression for the DR spectral-angular density	for the entire grating:

\begin{equation}
\frac{d^2 W_{GR}}{d\omega d\Omega}=\frac{d^2 W_{str}}{d\omega d\Omega}
F_N=\frac{d^2 W_{DR}}{d\omega d\Omega}F_{str}F_N,
\end{equation}

where

\begin{equation}
F_N=\left\vert\sum^N_{k=1}\exp[-(k-1)(\alpha_0+i\phi_0)]\right\vert^2=
\left\vert\frac{1-C^{N}}{1-C}\right\vert^2.
\end{equation}
Here $C=\exp(-\alpha_0-i\phi_0)$

After simple mathematical transformations, expression (24) may be written
in the following manner:

\begin{equation}
F_N=\exp\left[-(N-1)\alpha_0\right]\left\{\frac{\sin^2(\frac{N\phi_0}{2})+
\sinh^2(\frac{N\alpha_0}{2})}{\sin^2(\frac{\phi_0}{2})+
\sinh^2(\frac{\alpha_0}{2})}\right\}
\end{equation}

One should notice that the structure of formula (25) is identical to
that of a similar expression for the resonant transition radiation from $N$ layers taking
into account the radiation absorption in every layer.

First we shall consider a particular case corresponding to the Smith-Purcell geometry
$(\Theta_0=0)$. It is obvious that the decay factor $\alpha_0$ determied by
expression (20) is equal to zero, therefore, expression (25) can be rewritten in
a well-known form

\begin{equation}
F_N=\frac{\sin^2(\frac{N\phi_0}{2})}{\sin^2(\frac{\phi_0}{2})}
\end{equation}

When $N\rightarrow\infty$,  expression (26) transforms into an ordinary $\delta$-
function

\begin{equation}
F_N=2\pi N\delta(\phi_0-2k\pi)
\end{equation}
where $k$ is the diffraction order.

The presence of $\delta$-function is an indication of the existance of the monochromatic maxima in the RDR spectrum.
However, the use of the $\delta-$ function for real gratings
where the number of elements is limited is not always justified. Therefore, we shall further use the exact
formulas (25) and (26).

Figure 4 depicts the RDR spectral distribution for the Smith-Purcell geometry.
The calculation has been carried out for $a=\frac{d}{2}$
when the intensity reaches its maximum value [2]. The peak position in the spectrum
is determined by the phase relation (the resonance condition)

\begin{equation}
\phi_0=2k\pi,
\end{equation}

which leads to a well-known formula of Smith-Purcell

\begin{equation}
\lambda_k=\frac{d\left(\cos\Theta_y-\frac{1}{\beta}\right)}{k}
\end{equation}

As is seen from the figure, the even orders are absent.
This is explained by the influence of the $F_{str}$
factor that is equal to zero at

\begin{equation}
\phi=m\pi,
\end{equation}

where $m$ is the integer (see formula (11)).

Substituting (29) in (5), we can write expresion (28) in the following form:

\begin{equation}
k=\frac{d}{a}m=2m
\end{equation}

Thus, in the case under consideration ($a=\frac{d}{2}$) the even diffraction orders are
forbidden. For illustration, figure 5 shows the dependence of $F_{str}$ on the photon
energy which was calculated for the same conditions as in Fig.4.

Figure 6 shows the maximum DR yield dependence on
the ratio $\frac{a}{d}$ for four radiation orders .

As was shown above, for the first diffraction order the intensity reaches
its maximum value at $\frac{a}{d}=0.5$, whereas for the second order $-$ at $\frac{a}{d}
=0.25$ and $0.75$.

Let us consider the case of a tilted grating. One should note that in the DR spectrum the
quasimonochromatic peaks can be observed at small tilt angles of the grating.
For the tilted single strip the spectrum calculated according to (10) is shown in
Fig.7. The strip parameters and geometry are indicated in the figure caption.
As in the case of a semi-infinite screen one can observe an exponentially decreasing
spectrum.

Figure 8 depicts the RDR spectrum from a tilted grating
calculated at $\Theta_0=1.9^0$. Unlike the spectrum from a single strip
calculated for the same initial conditions, here one can observe
the first order quasimonochromatic radiation maximum with a finite full width
at half maximum (FWHM) together with the continuous background.

The position of the quasimonochromatic maxima in the RDR spectrum is determined by the
resonance condition (28) where the $\phi_0$ phase is taken according to formula
(19). We will illustrate this fact in the following way. Let us rewrite expression
(25) in the form

\begin{equation}
F_N=\frac{1-2\exp\{-(N-1)\alpha_0\}\cos\{(N-1)\phi_0\}+\exp\{-2(N-1)\alpha_0\}}{1-2\exp
\{-\alpha_0\}\cos\phi_0+\exp\{-2\alpha_0\}}
\end{equation}

In extreme case, when $N\rightarrow\infty$, instead of (32) we have

\begin{equation}
F_{\infty}=\frac{1}{1-2e^{-\alpha_0}\cos\phi_0+e^{-2\alpha_0}}
\end{equation}

It is apparent that the expression obtained reaches its maximum value when the following
conditions are fulfilled:

\begin{equation}
\phi_0=2k\pi \hspace{10mm}(\cos\phi_0=1)
\end{equation}

In this case:

\begin{equation}
F_{\infty}=\frac{1}{(1-e^{-\alpha_0})^2}
\end{equation}

At small values, $\alpha_0<<1$, that correspond to small tilt angles of
the grating, $\Theta_0<<1$, from (35) we have

\begin{equation}
F_{\infty}\cong\frac{1}{\alpha_0^2}
\end{equation}

Thus, the pronounced quasimonochromatic maxima in the DR spectrum can be observed
at the sliding incidence angles of the particle beam with respect to the grating.
For the angle $\Theta_0\neq 0$, instead of Smith-Purcell conditions we have the
following relation between the quasimonochromatic maximum position, period $d$,
grating tilt angle $\Theta_0$ and observation angle $\Theta_y$:

\begin{equation}
\lambda_k=\frac{d\left[\cos(\Theta_y-\Theta_0)-\frac{\cos\Theta_0}{\beta}\right]}
{k},
\end{equation}

where $k$ is the integer.

Figure 9a shows the dependence of different - order maximum positions on the
target orientation angle at the observation angle $\Theta_y=4.5^0$.

One can notice that for negative values of $\Theta_0$ (i.e. for the geometry
where the beam "reflected" by the grating is directed to the opposite side from
the detector) the spectral maxima are shifted in the soft part with respect to
the SPE spectrum ($\Theta_0=0$). When the tilt angle increases, the spectral
maxima are shifted to the high energy part.

A similar peak shift of the parametric X-ray radiation has been registered in the experiment [11] with rotating a
crystallic target. Actually,
for both the RDR and PXR the peak position in the spectrum is determined by the resonance
condition only and does not depend on the radiation mechanism.

The experiments on the PXR studies often measure the so-called
theta-scan, i.e., the dependence of the radiation yield for a fixed observation
angle and electron energy on the target orientation angle.
Figure 9b presents this dependence calculated for the RDR.
In this case the grating tilt angle $\Theta_0$ is varied with respect to
the electron beam.

A recent experiment [12] measured the same
dependence for the Smith-Purcell effect for the grating made as a periodically
deformed continuous surface. The authors of
the quoted paper obtained the dependence with clear maxima.
This dependence shape is quite close to the one presented in Fig.9b.

Figure 10a presents the dependence of the full width at half maximum
($\Delta E$) on the number of grating elements for the first diffraction order
and tilt angle $\Theta_0=1^0$ (solid line).

It also shows a similar dependence for the SPE which is well
approximated by the $\frac{1}{N}$ dependence (dotted line). As follows from the figure,
for $\Theta_0=1^0$ the resulting curve is well approximated by the $\frac{C_1}{N}$
formula where $C_1=1.8$.

When the number of periods increases, the RDR intensity also increases reaching
the $0.95I_{\infty}$ level for $N=70$ (see Fig.10b).

Let us estimate the effective grating length (the number of periods) with which
the passing particle field interacts

\begin{equation}
N_{eff}=\frac{\gamma\lambda}{d\sin\Theta_0}
\end{equation}

In the case considered $N_{eff}=60$, which is quite close to the grating
length providing a $95\%$ intensity level.

When the particle moves close to the grating of a limited length (with the
number of periods $N$), one can derive a similar caracteristics for
the Lorentz-factor

\begin{equation}
\gamma_{eff}=N \frac{d\sin\Theta_0}{\lambda}
\end{equation}

Figure 11 shows the dependence of $I_{max}$ on the particle energy. As
follows from the figure the simple estimation (39) is a good characteristics of the RDR
process too.

\hspace{8cm}\section*{Summary}

A simple model for calculating the RDR characteristics from a tilted grating has
been suggested. It has been shown that quasimonochromatic
maxima appear in the RDR spectrum, and their characteristics (full width, intensity) are determined primarily by
the angle between the grating plane and electron pulse.

Let us estimate the initial beam divergence effect for the SP geometry. It was noticed in [4] that for the angles
$\Theta_H<<1$ ($\Theta_H$ is the angle between the electron momentum and grating axis in the
horizontal plane) the peak shift can be described in the following way:

\begin{equation}
\lambda_k = \frac{d}{\cos\Theta_H}\cdot\frac{\cos\Theta_y-\frac{1}{\beta}}{k}
\simeq \lambda_k^0 \left (1+\frac{\Theta_H^2}{2}\right)
\end{equation}

Thus,

\begin{equation}
\frac{\Delta\lambda_H}{\lambda} = \frac{\Theta_H^2}{2},
\end{equation}

therefore, the peak broadening related to the beam devergence in horizontal plane can be neglected if

\begin{equation}
\frac{\Theta_H^2}{2}<<\frac{1}{N},
\end{equation}

which is generally fulfilled.

In order to obtain a formula, analogous to (41) and characterizing the beam broadening due to the divergence in the
vertical plane, we shall take into account that the observation angle

\begin{equation}
\Theta_y - \Theta_0 = \Theta_D = const,
\end{equation}

where the angle $\Theta_D$ is measured from the grating plane.

Let us denote the angle $\Theta_0$ as $\Theta_V$ ($\Theta_V$ is the angle between the electron pulse and grating axis
in the vertical plane). Then for $\Theta_V<<1$ we have

\begin{equation}
\lambda_k = \frac{d}{k}\left [ \cos \Theta_D-\frac{1-\frac{\Theta_V^2}{2}}{\beta}\right]
\end{equation}

It follows from (44) that the SPE peak broadening is determined by the observation angle $\Theta_D$

\begin{equation}
\frac{\Delta\lambda_V}{\lambda} \simeq -\frac{\Theta_V^2}{4\sin^2\frac{\Theta_D}{2}}
\end{equation}

For the small observation angles $\Theta_D$ the peak broadening $\frac{\Delta\lambda_V}{\lambda}$
can be significant and exceed the "natural" peak width $\frac{1}{N}$.
Thus, changing the SPR peak width, for the small angles $\Theta_D$, in principle, one can
determine the vertical beam divergence.

From (45) one can estimate the sensitivity of the method:

\begin{equation}
\Theta_V\geq\frac{2\sin\frac{\Theta_D}{2}}{\sqrt{N}}
\end{equation}

Since the latter expression does not depend on the wavelength,
investigation of the shape line can be carried out in the optical region. For example, for
$\Theta_D = 2^0$ and $N=100$ $\Theta_V\sim 3.5\cdot 10^{-3}$. The estimation obtained does not
depend on $\gamma$ (if $\gamma>>1$).

The suggested technique for the beam divergence determination can be used in accelerators
with $\gamma \leq 100$, because, in this case, the well-known methods based on either
transition radiation or synchrotron radiation exhibit a sensitivity controlled
by the characteristic radiation angle $\gamma^{-1}$.

The dependence of the peak position in the RDR spectrum on the grating tilt angle $\Theta_0$ can be used
to determine the electron bunch length $l_e$. In [6,13] the authors suggested to measure
the coherent SPR yield at different observation angles. In the wavelength region of $\lambda \sim l_e$
one will observe conversion from an ordinary SPR to coherent SPR, i.e., the radiation intensity
will be changed by approximately $N_e$ times ($N_e$ is the number of electrons in a bunch).

Figure 12a shows the dependence of the RDR wavelength on the grating tilt angle for the
fixed observation angle $\Theta_y$, and a similar dependence for the peak width is presented
in Fig. 12b. The calculations have been carried out for the following conditions:

$$
\gamma = 100,
\Theta_y = 4.5^0,
d = 20mm,
a = 10mm.
$$

As follows from the figure, in the wavelength region $\lambda=0.15 \div 0.8mm (\Theta_0=8^0\div 35^0)$
a practically constant intensity of the ordinary RDR is observed. If we measure the RDR intensity
for the electron bunch
with the length $l_e\sim 0.5mm$, then for the grating tilt angles indicated one could research the conversion to
the "coherent
RDR mode" in detail, which would allow one to determine both the everage bunch length and the profile electron
distribution in a bunch. The suggested technique is related to the grating rotation for a fixed detector position,
whereas in [6,13] it was sugested to move the detector, which is not always convinient (in particular,
if the detecting system contains a monochromator).

Depending on the bunch length $l_e$, the region of the investegated wavelengths can be easily changed by
choosing a proper grating period and observation angle.

\begin{center}
\section*{Acknowlegements}
\end{center}

The present work has been carried out under a partial support of the
Russian Basic Research Fund (grants No. 98-02-17994 and No. 99-02-16884).

\begin{center}
\hspace{6cm}\section*{References}
\end{center}

[1]  O.Haeberle, P.Rullhusen et al. Phys.Rev. 49 (1994) 3340.

[2]  A.P.Potylitsyn, Phys.Lett. A 238 (1998) 112-116.

[3]  K.J.Woods, J.E.Walsh, and R.E.Stoner, Phys.Rev. E 49 (1994) 3340.

[4]  O.Haeberle, P.Rullhusen et al. Phys.Rev. E 55 (1997) 4675.

[5]  Y.Shibata, S.Hasebe, K,Ishi et al. Phys.Rev. E 57 (1998) 1063.

[6]  M.C.Lampel NIM, A 385 (1997) 19-25.

[7]  A.P.Kazantsev and G.I.Surdutovich, Sov.Phys.Dokl. 7 (1963) 990.

[8]  A.P.Potylitsyn, NIM B 145 (1998) 169-179.

[9]  A.P.Potylitsyn, NIM B 145 (1998) 60-66.

[10] X.Artru, G.B.Yodh, and G.Mennessier, Phys.Rev. D 12 (1975)1289.

[11] Yu.N.Adishev, V.A.Versilov et al., NIM B 44 (1989) 130-136.

[12] G.Kube et al., Phys.Rev. E (to be published).

[13] Dinh C. Nguyen, NIM A 393 (1997) 514-518.

\begin{center}
\hspace{6cm}\section*{Figure captions}
\end{center}

Fig.1. Geometry of diffraction radiation (DR) from a single strip; $h$ is the
impact parameter, $\Theta_y$ is the observation angle, $\Theta_0$ is the
strip tilt angle, and $a$ is the strip width.

\vspace{0.5cm}
Fig.2. Orientation dependence for a single strip with the widths
a=0.1mm (a), 1mm (b), 10mm (c), and a semi-infinite plane (d).

\vspace{0.5cm}
Fig.3. Geometry of diffraction radiation from a tilted grating;
$h$ is the impact parameter, $a$ is the sprip width, $d$ is the
grating period, and $\Theta_0$ is the grating tilt angle.

\vspace{0.5cm}
Fig.4. Smith-Purcell effect spectrum. The initial conditions used are
$\Theta_y=4.5^0$, $\Theta_x=0$, $\Theta_0=0$, $a=0.2mm$, $d=0.4mm$,
$h=0.1mm$, $\gamma=1000$, $N=50$; $k$ is the diffraction order.

\vspace{0.5cm}
Fig.5. Dependence of $F_{STR}$ on energy.

\vspace{0.5cm}
Fig.6. Dependence of the first (1), second (2), third (3) and fourth (4) order maximum intensity
on the ratio $\frac{a}{d}$. The initial conditions used are
$\Theta_y=4.5^0$, $\Theta_x=0$, $\Theta_0=0$, $h=0.1mm$, $\gamma=1000$,
$N=50$, $\lambda=0.4\mu m$.

\vspace{0.5cm}
Fig.7. Diffraction radiation spectrum from a tilted strip. The initial
conditions used are $\Theta_y=4.5^0$, $\Theta_x=0$, $\Theta_0=1.9^0$,
$a=0.2mm$, $a_1=0.1mm$ (see Fig.1.), $\gamma=1000$.

\vspace{0.5cm}
Fig.8. Diffraction radiation spectrum from a tilted grating. The initial
conditions used are $\Theta_y=4.5^0$, $\Theta_x=0$, $\Theta_0=1.9^0$,
$a=0.2mm$, $h=0.1mm$ (see Fig.3.), $\gamma=1000$.

\vspace{0.5cm}
Fig.9. a: Dependence of the first (1), second (2), third (3) and fourth (4)
order peak positions ($\Theta_y=4.5^0$, $d=0.4mm$) on the tilt angle.

\hspace{1.2cm} b: Dependence of the
first DR order yield ($\lambda=1233nm$) on the tilt angle.

\vspace{0.5cm}
Fig.10. a: Dependence of the first order FWHM for SPE ($\Theta_0=0$)
(solid line), $\Theta_0=1^0$ (dash-dotted line) on the number of
elements and  dependence of the $\frac{1}{N}$ (dashed line). The initial
conditions used are $\Theta_y=4.5^0$, $\Theta_x=0$,$a=0.2mm$, $h=0.1mm$,
$d=0.4mm$ (see Fig.1.), $\gamma=1000$.

\hspace{1.3cm} b: Dependence the first order maximum intensity	on the number of
elements calculated for the same initial conditions.

\vspace{0.5cm}
Fig.11. a:  Dependence of the first order FWHM (dash-dotted line) on the $N_{eff}$
and dependence of $\frac{1}{N_{eff}}$ (solid line). The initial
conditions used are $\Theta_y=4.5^0$, $\Theta_x=0$, $\Theta_0=1^0$,
$a=0.2mm$, $h=0.1mm$, $d=0.4mm$ (see Fig.1.), $\gamma=1000$.

\hspace{1.3cm} b: Dependence of the first order maximum
intensity on the $N_{eff}$ calculated for the same initial conditions.

\vspace{0.5cm}
Fig.12. a: Dependences of the RDR peak intensity on the grating tilt angle
(radiation wavelength) for $\gamma = 100, \Theta_y=4.5^0, d=20mm, a=10mm$.

\hspace{1.3cm} b: Dependence of the FWHM on the grating tilt angle (radiation
wavelength)  for the same initial conditions.

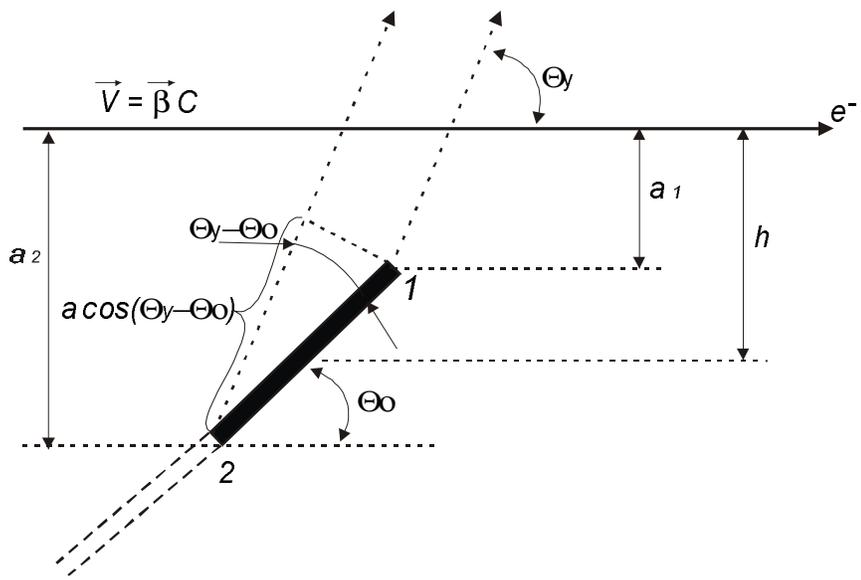
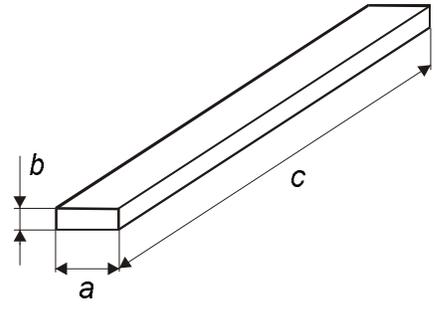
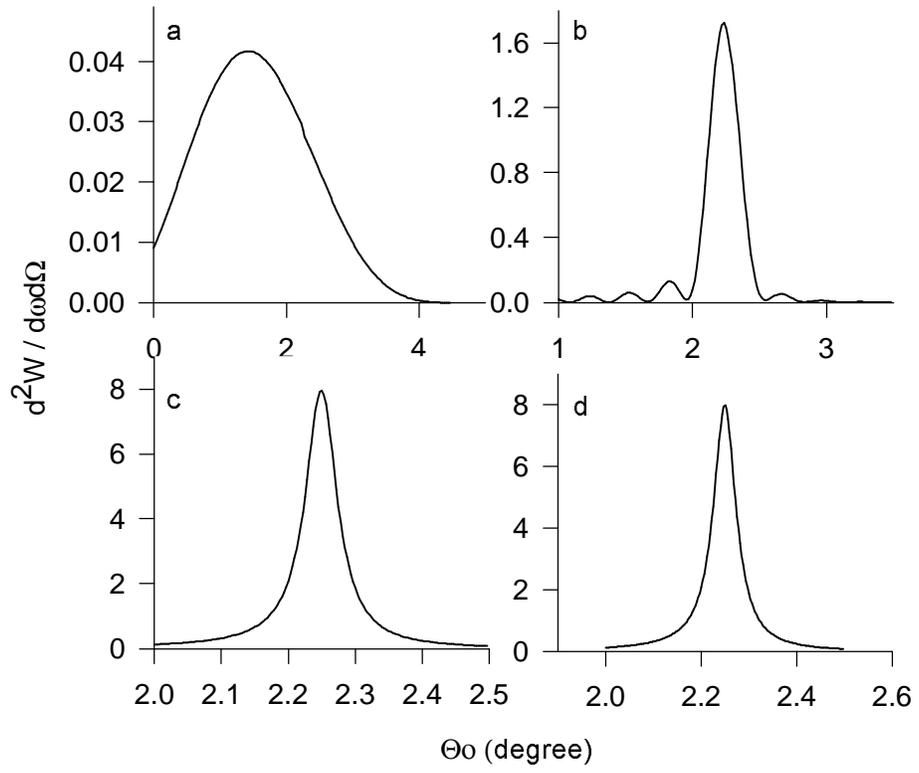
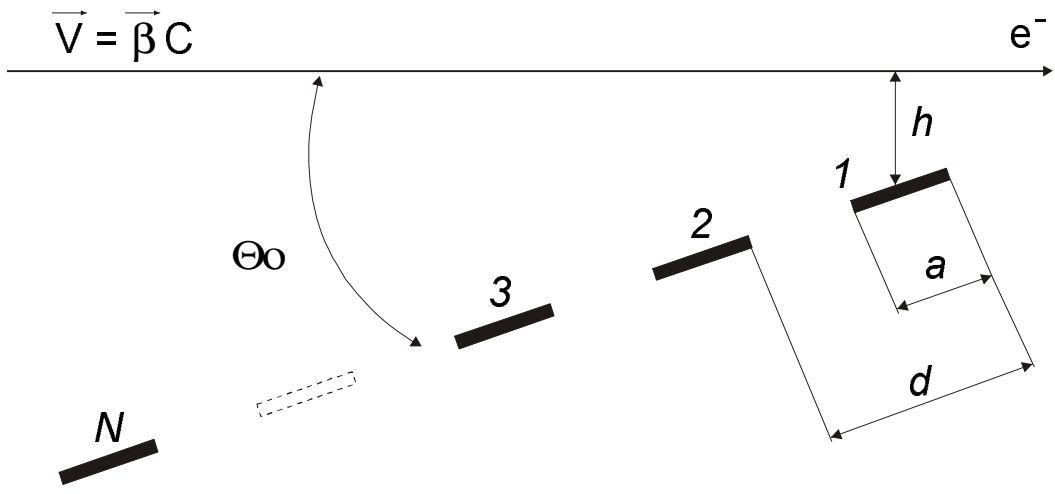
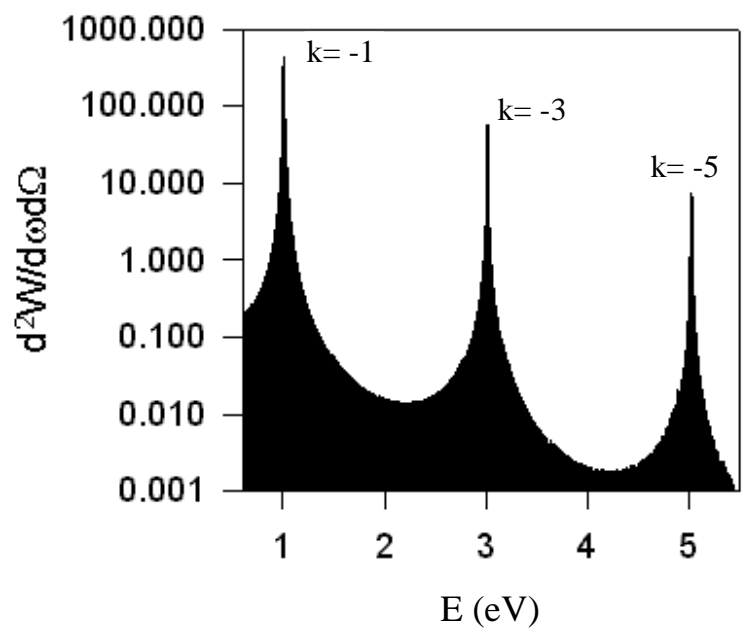
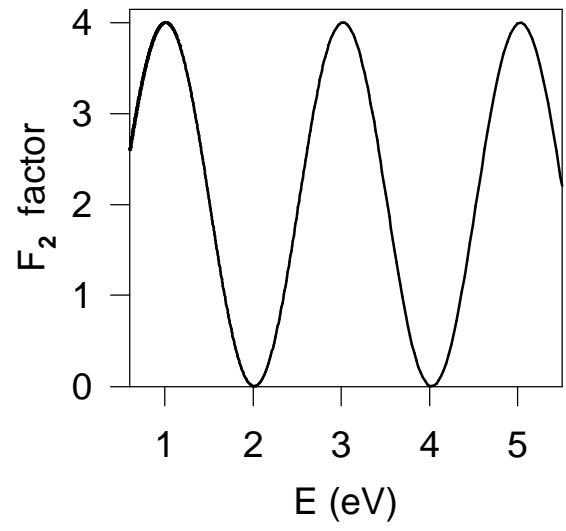
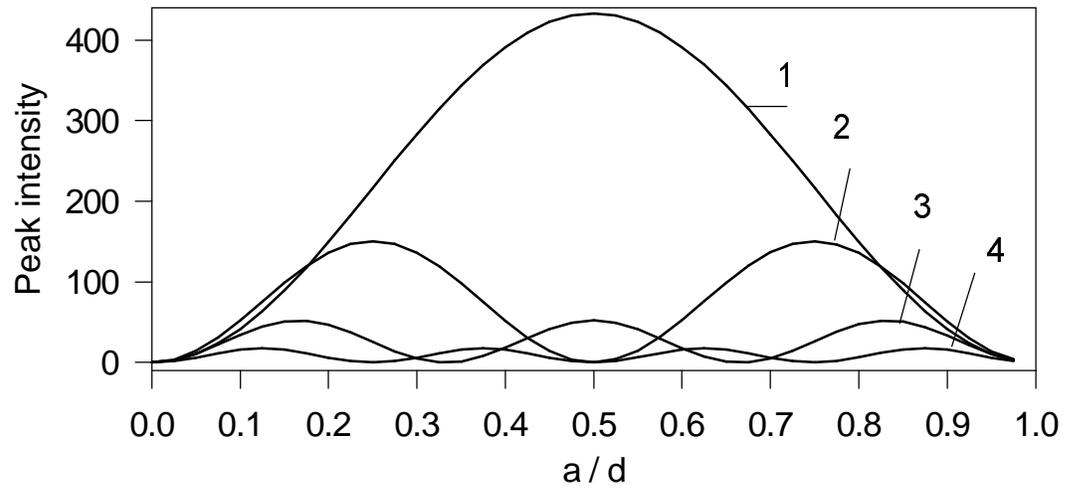
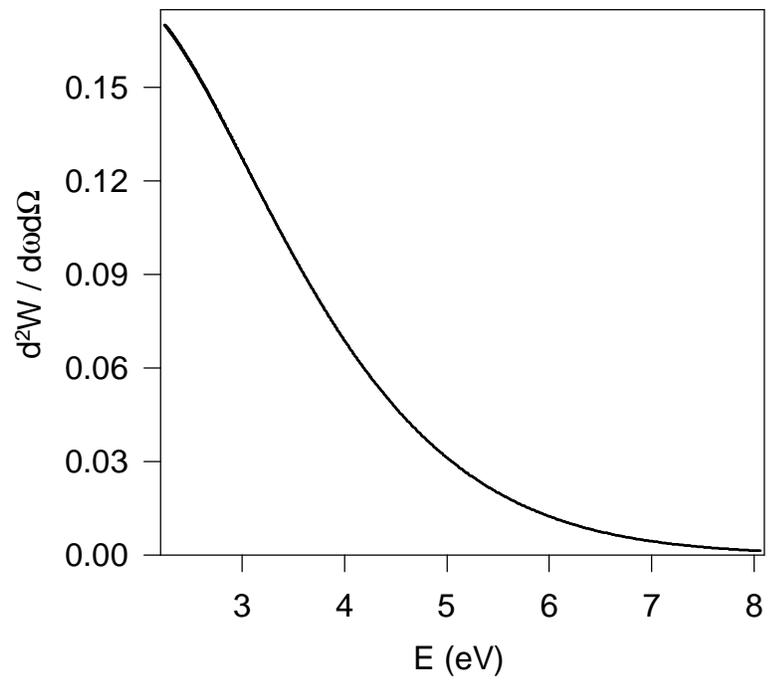
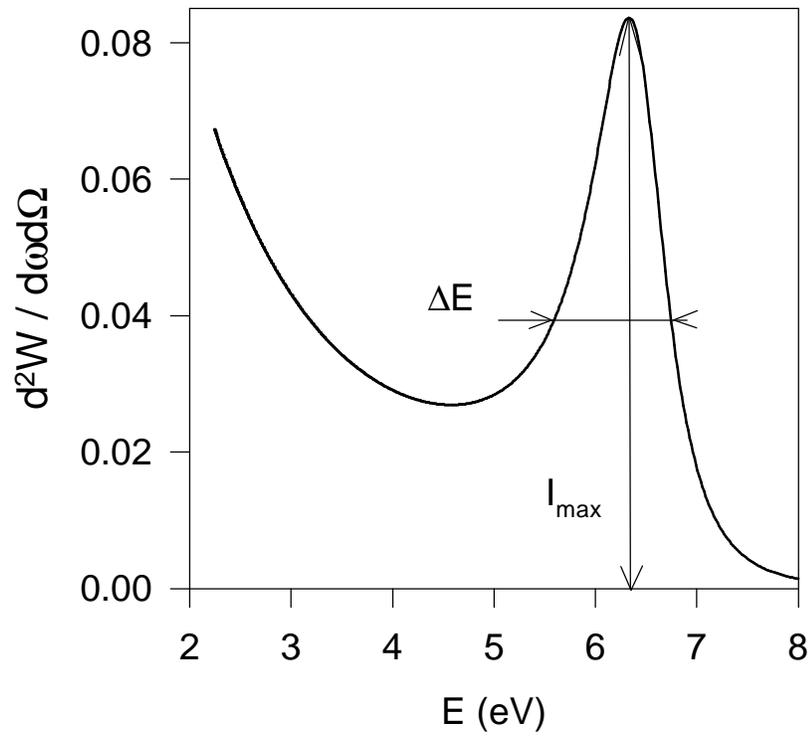
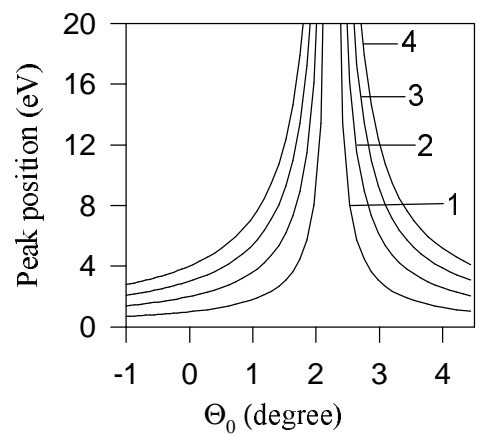
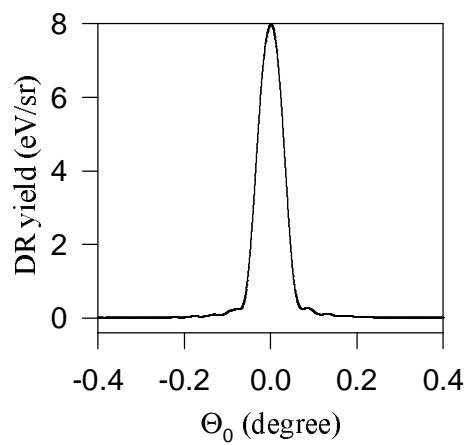
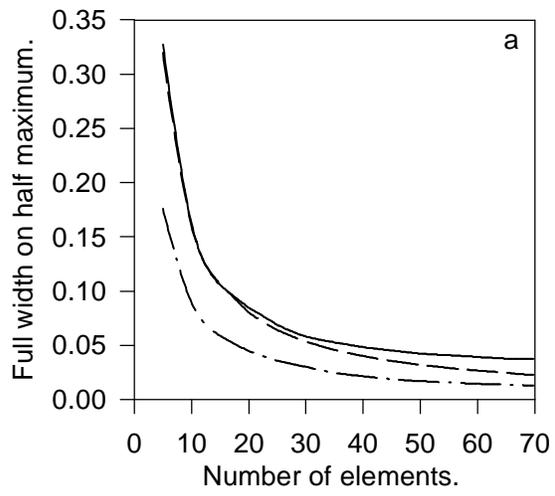
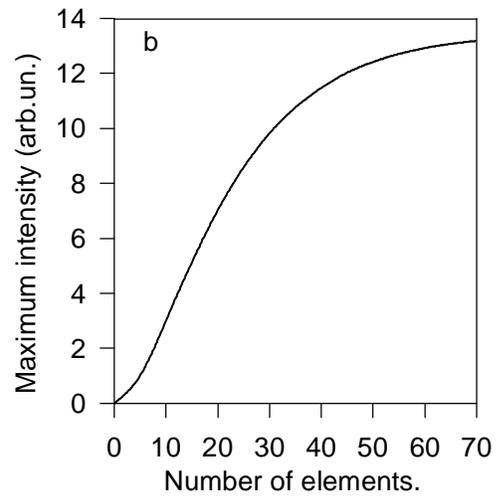
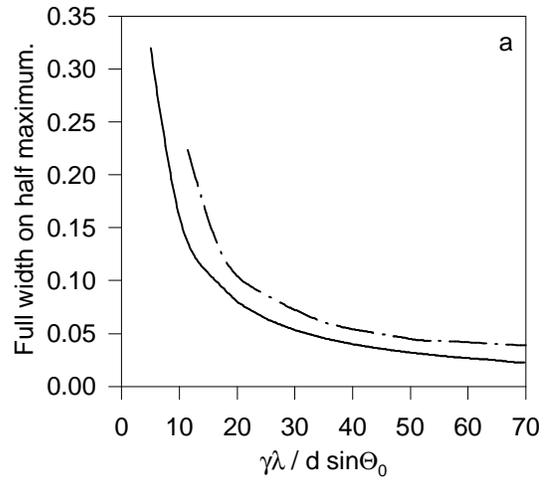
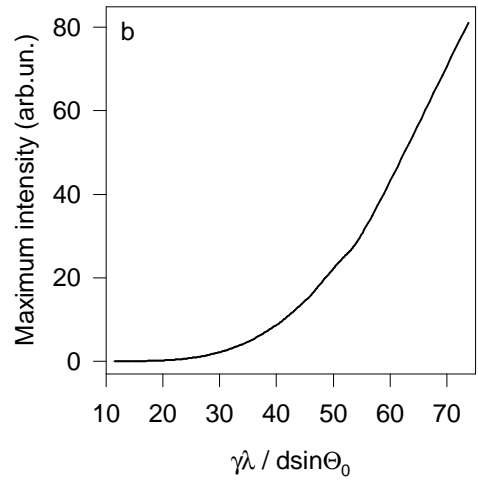
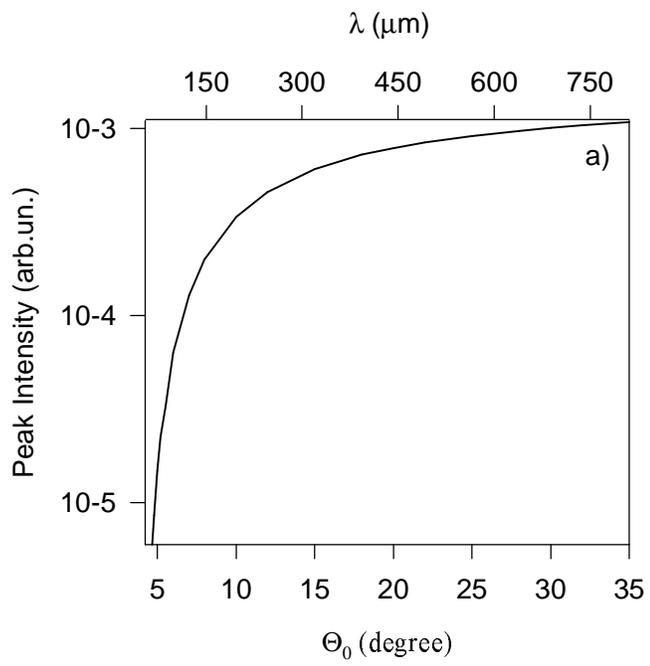
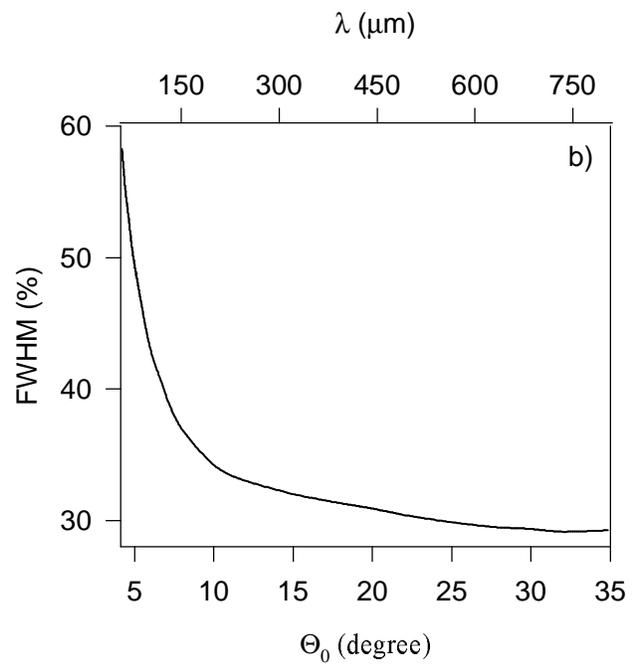
\end{document}